\documentstyle[aps,epsf]{revtex}
\begin{document}
\draft
\title{Two-dimensional $S=1/2$ antiferromagnet on a plaquette lattice}
\author{S.V. Meshkov\cite{byline1}}
\address{Laboratoire des Verres\\ Universit\'e de Montpellier II, CNRS UMR 5587\\
Place E.Bataillon, 34095 Montpellier, France}
\author{D. Foerster\cite{byline2}}
\address{Centre de Physique Th\'eorique et de Mod\'elisation de Bordeaux\\
Universit\'e de Bordeaux I, CNRS URA 1537\\
rue de Solarium, 33175 Gradignan, France}
\maketitle
\begin{abstract}

We consider a simplified model of the magnetic structure of the
two-dimensional compound $CaV_4O_9$ in terms of interacting square plaquettes
of spins with two distinct antiferromagnetic exchange constants.  We analyze
the competition between two types of singlet ground states and the Neel ordered
one in terms of respectively, numerical cluster expansion and nonlinear spin
wave theory.  The resulting phase diagram agrees well with known Quantum Monte
Carlo results and suggests a first order transition between ordered and
singlet ground states as a function of the exchange constants.  
\end{abstract}
\pacs{PACS numbers: 75.10.-b, 75.10.Jm}

The recent experimental observation of a spin gap in the layered $S=1/2$
antiferromagnet $CaV_4O_9$ \cite{r010} has opened a new and interesting
perspective in two dimensional magnetism. We discuss the simplest model of
the undoped structure that consists of a square lattice of elementary
squares or ''plaquettes'', which we will refer to as CAVO lattice. The
magnetic exchange energy within the plaquettes ($J_0$) and between the
plaquettes ($J_1$) is given by 
\begin{equation}
\widehat{H}=J_0\sum_{\Box}{\bf S}_i{\bf S}_j+J_1\sum_{-}{\bf S}_i{\bf S}_j
\label{e10}
\end{equation}
Here $ij$ represent nearest neighbors on edges of a plaquette and between
adjacent plaquettes, respectively. Additional (frustrating) couplings are
ignored, although they are believed to be necessary for quantitative
agreement with experiments.

The nature of the ground state is easy to understand in two limits \cite{r020}. 
In the limit of $J_1\ll J_0$ the plaquettes form resonating valence bond
type singlet states, with an energy of $-\frac 12J_0$ per plaquette, and
weak bonds $J_1$ serve as a perturbation. In the opposite limit of $J_0\ll
J_1$ the interplaquette connections form singlets of energy $-\frac 38J_1$
per dimer that are weakly interacting via plaquettes. This construction is
qualitatively symmetric, but the plaquettes are somewhat ``stronger'', so
that the critical ``equilibrium'' region is centered at $J_1\simeq\frac 43
J_0$. In this region, in addition to these two quantum singlet phases, the
antiferromagnetically ordered Neel state could also be competitive.

A variety of theoretical methods was applied to study the ground states of this
model as a function of $J_1/J_0$ but the results remain contradictory.  Our
purpose is to compare the energies of three candidates for the ground state of
the model (\ref{e10}) and estimate the regions of their stability in terms of
the ratio $J_1/J_0$. 
Unlike other approaches that attempted
to treat the system within a unified framework for all $J_1/J_0$, we choose
the most quantitatively reliable approach for each individual phase. For the
two singlet states we develop a numerical perturbation expansion (beyond the
second order \cite{r020}) in the coupling ratio $J_1/J_0$ or $J_0/J_1$,
whichever is smaller. The energy of the Neel state is estimated via the
nonlinear spin wave approximation, which gives a lower ground state energy
than the linear approximation reported in \cite{r030}. We also compare our
results with ground state energies obtained by direct numerical
diagonalisation of the Hamiltonian for different finite lattices of up to 24
spins and with frustrating/nonfrustrating boundary conditions.

\section{Cluster expansion for the ground state}

The idea of the cluster expansion is quite general and works for any model with
finite range interactions provided the unperturbed Hamiltonian is a sum
commuting blocks.
 For the model under consideration and in both limits
$J_1\ll J_0$ and $J_0\ll J_1$ the perturbation is a sum of exchange interactions
between nearest neighbors, and the zero approximation (the first or
second term in Eq.(\ref{e10}) respectively) is a sum over independent 
plaquettes or dimers.  Thus in any given order $n$ of the perturbation
parameter (the smaller of the coupling ratios) the total correction to the
ground state energy of an arbitrary lattice cluster $c$ can be reorganized into
a sum over connected graphs that mark the interactions that were used and the
unperturbed blocks that were touched:
\begin{equation}
E_c^{(n)}=\sum_gN_{c,g}\varepsilon_g(n)\,. \label{e20}
\end{equation}
Here $\varepsilon_g(n)$ stands for the contribution of the graph $g$ in the
order $n$ and $N_{c,g}$ for the number of ways the graph $g$ can be embedded in
the cluster $c$. The same relation holds for the entire infinite lattice with
properly normalized embedding numbers $N_\infty(g)$.

The detailed analysis is rather tedious (see e.g. similar analysis \cite
{r040} and references therein), and we only list the resulting sequences of
graphs and embedding numbers. To be specific, consider the $J_1$ expansion
about the plaquette state, with four graphs contributing up to the 5th order
\begin{equation}
1^{(II)}\;\begin{picture}(4,3.5)(0,0)
\put(1,1){\line(1,0){1}}
\thicklines
\put(0,0){\framebox(1,1)}\put(2,1){\framebox(1,1)}
\end{picture}
2^{(IV)}\;\begin{picture}(6,3.5)(0,0)
\put(1,1){\line(1,0){1}}\put(3,2){\line(1,0){1}}
\thicklines
\put(0,0){\framebox(1,1)}\put(2,1){\framebox(1,1)}\put(4,2){\framebox(1,1)}
\end{picture}
3^{(IV)}\;\begin{picture}(5,3.5)(0,0)
\put(1,2){\line(1,0){1}}\put(3,1){\line(0,1){1}}
\thicklines
\put(0,1){\framebox(1,1)}\put(2,2){\framebox(1,1)}\put(3,0){\framebox(1,1)}
\end{picture}
4^{(IV)}\;\begin{picture}(5,3.5)(0,0)
\put(1,2){\line(1,0){1}}\put(2,0){\line(1,0){1}}\put(3,1){\line(0,1){1}}
\put(1,0){\line(0,1){1}}
\thicklines
\put(0,1){\framebox(1,1)}\put(2,2){\framebox(1,1)}\put(3,0){\framebox(1,1)}
\put(1,-1){\framebox(1,1)}
\end{picture}
\label{e30}
\end{equation}

We draw the graphs as being embedded in the CAVO lattice, marking the $J_0$
connections by bold lines. The Roman superscript of a graph denotes the lowest
order of perturbation in which it contributes to Eq.(\ref{e20}). A connected
graph contributes to order $n$ if $n$ interactions can be placed on the links
of the graph (one or more per link) such that (i) each block is touched at least
twice (otherwise the block cannot return to its singlet ground state), (ii) no
two parts of the graph are connected by only one interaction (it would vanish
by spin symmetry).  Once the contributions $\varepsilon_g(n)$ of all necessary
graphs are known, the $n$-th order corrections $E_c^{(n)}$ to the ground state
energy of an arbitrary lattice cluster $c$ can be calculated using the
embedding numbers $N_{c,g}$.  One can, however, invert the procedure and
recover the contributions of the graphs by solving the system of linear
equations (\ref{e20}) for $\varepsilon_g(n)$ in terms of the $n$-th order
corrections $E_c^{(n)}$ of an appropriately selected set of small clusters $c$.

The most economic or ''optimal'' set of clusters is obtained when each
cluster embeds some graph in the list of graphs exactly once.
Note that the graphs are topological 
entities, so that the choice of optimal clusters is generally not unique.
Once a choice has been made, there is a one-to-one
correspondence between the graphs contributing to a given
order and the finite clusters carrying the information on their contributions. 
Therefore we can use the same pictures and labels for optimal clusters as
for graphs.
For the plaquette expansion the embedding numbers $N_{c,g}$ of all
graphs of (\ref{e30}) in all clusters of (\ref{e30}) are 
\[
{\tighten
N_{c,g}^{\rm pla}=\left(
\begin{array}{cccc}
1&0&0&0\\ 
2&1&0&0\\ 
2&0&1&0\\ 
4&0&4&1  
\end{array}
\right)
}
\]
with rows of the matrix corresponding to clusters and columns corresponding to
graphs.  

The coefficients $E_c^{(n)}$ are easily extracted from a polynomial fit of high
precision ground state energies of the clusters in the list at several values of
the coupling ratio.  Having solved for $\varepsilon_g(n)$ from Eq.(\ref{e20})
we calculate (to the same 5th order) the ground state energy per site of an
infinite lattice with embedding constants $N_{\infty,g}^{{\rm pla}}=\frac
14\left(2\,2\,4\,1\right)$ and obtain 
\begin{equation}
E_{{\rm pla}}(J_0,J_1) = 
J_0\left[ -\frac 12-\frac{43}{1152}\left(
\frac{J_1}{J_0}\right)^2-0.00723\left(\frac{J_1}{J_0}\right)^3
-0.00308\left(
\frac{J_1}{J_0}\right)^4-0.0022\left(\frac{J_1}{J_0}\right)^5\right]  +... 
\label{e40} 
\end{equation}
So we find 5 orders of the perturbation series for the {\em infinite}
lattice by diagonalizing only 4 {\em finite} and relatively small clusters
of up to 16 spins by use of conventional Lanczos algorithms.

In the case of the dimer expansion in the small parameter $J_0/J_1$ 
the unperturbed blocks are smaller (2 sites), so that we can reach 7th
order of perturbation with clusters not exceeding 12 sites. The list of
graphs/clusters contains 13 entries 
\begin{eqnarray}
&
1^{(II)}\;\begin{picture}(3.5,3)(0,0)
\thicklines
\put(0,0){\line(1,0){1}}
\thinlines
\put(1,0){\line(1,0){1}}\put(0,0){\line(0,1){1}}
\end{picture}
2^{(IV)}\;\begin{picture}(3.5,3)(0,0)
\thicklines
\put(0,1){\line(0,1){1}}\put(0,0){\line(1,0){1}}
\thinlines
\put(0,2){\line(1,0){1}}\put(1,0){\line(1,0){1}}\put(0,0){\line(0,1){1}}
\end{picture}
3^{(IV)}\,\begin{picture}(4.5,3)(0,0)
\thicklines
\put(1,0){\line(0,1){1}}\put(1,1){\line(1,0){1}}
\thinlines
\put(0,0){\line(1,0){1}}\put(2,1){\line(1,0){1}}\put(1,1){\line(0,1){1}}
\end{picture}
4^{(IV)}\;\begin{picture}(3.5,3)(0,0)
\thicklines
\put(0,1){\line(0,1){1}}\put(2,0){\line(0,1){1}}\put(1,2){\line(1,0){1}}
\put(0,0){\line(1,0){1}}
\thinlines
\put(0,2){\line(1,0){1}}\put(1,0){\line(1,0){1}}\put(2,1){\line(0,1){1}}
\put(0,0){\line(0,1){1}}
\end{picture}
5^{(IV)}\,\begin{picture}(4.5,3)(0,0)
\thicklines
\put(1,0){\framebox(1,1)}
\thinlines
\put(0,0){\line(1,0){1}}\put(2,1){\line(1,0){1}}\put(1,1){\line(0,1){1}}
\put(2,0){\line(0,-1){1}}
\end{picture}
6^{(VI)}\;\begin{picture}(3.5,3)(0,0)
\thicklines
\put(0,1){\line(0,1){1}}\put(1,2){\line(1,0){1}}\put(0,0){\line(1,0){1}}
\thinlines
\put(0,2){\line(1,0){1}}\put(1,0){\line(1,0){1}}\put(2,1){\line(0,1){1}}
\put(0,0){\line(0,1){1}}
\end{picture}
7^{(VI)}\,\begin{picture}(4.5,3)(0,0)
\thicklines
\put(1,0){\line(1,0){1}}\put(1,0){\line(0,1){1}}\put(1,1){\line(1,0){1}}
\thinlines
\put(0,0){\line(1,0){1}}\put(2,1){\line(1,0){1}}\put(1,1){\line(0,1){1}}
\put(2,0){\line(0,-1){1}}
\end{picture}
\nonumber\\
&
8^{(VI)}\,\begin{picture}(4.5,3)(0,0)
\thicklines
\put(1,0){\line(0,1){1}}\put(1,1){\line(1,0){1}}\put(3,1){\line(0,1){1}}
\thinlines
\put(0,0){\line(1,0){1}}\put(2,1){\line(1,0){1}}\put(1,1){\line(0,1){1}}
\put(3,2){\line(0,1){1}}
\end{picture}
9^{(VI)}\,\begin{picture}(3.5,3)(0,0)
\put(0,0){\line(1,0){1}}\put(2,1){\line(1,0){1}}\put(1,1){\line(0,1){1}}
\put(1,3){\line(1,0){1}}
\thicklines
\put(1,0){\line(0,1){1}}\put(1,1){\line(1,0){1}}\put(1,2){\line(0,1){1}}
\end{picture}
10^{(VI)}\;\begin{picture}(4.5,3)(0,0)
\put(0,2){\line(1,0){1}}\put(1,0){\line(1,0){1}}\put(2,1){\line(0,1){1}}
\put(0,0){\line(0,1){1}}\put(3,0){\line(0,-1){1}}
\thicklines
\put(0,1){\line(0,1){1}}\put(2,0){\line(0,1){1}}\put(1,2){\line(1,0){1}}
\put(0,0){\line(1,0){1}}\put(2,0){\line(1,0){1}}
\end{picture}
11^{(VI)}\,\begin{picture}(4.5,3)(0,0)
\put(0,0){\line(1,0){1}}\put(2,1){\line(1,0){1}}\put(1,1){\line(0,1){1}}
\put(2,0){\line(0,-1){1}}\put(3,2){\line(0,1){1}}
\thicklines
\put(1,0){\framebox(1,1)}\put(3,1){\line(0,1){1}}
\end{picture}
12^{(VI)}\;\begin{picture}(4.5,3)(0,0)
\put(0,2){\line(1,0){1}}\put(1,0){\line(1,0){1}}\put(2,1){\line(0,1){1}}
\put(0,0){\line(0,1){1}}\put(3,0){\line(0,-1){1}}\put(3,1){\line(1,0){1}}
\thicklines
\put(0,1){\line(0,1){1}}\put(1,2){\line(1,0){1}}
\put(0,0){\line(1,0){1}}\put(2,0){\line(1,0){1}}
\put(3,0){\line(0,1){1}}\put(2,1){\line(1,0){1}}
\end{picture}
13^{(VII)}\;\begin{picture}(3.5,3)(0,0)
\put(0,2){\line(1,0){1}}\put(1,0){\line(1,0){1}}\put(2,1){\line(0,1){1}}
\put(0,0){\line(0,1){1}}\put(3,0){\line(0,-1){1}}\put(3,1){\line(1,0){1}}
\thicklines
\put(0,1){\line(0,1){1}}\put(2,0){\line(0,1){1}}\put(1,2){\line(1,0){1}}
\put(0,0){\line(1,0){1}}\put(2,0){\line(1,0){1}}
\put(3,0){\line(0,1){1}}\put(2,1){\line(1,0){1}}
\end{picture}
\label{e50}
\end{eqnarray}
in the same notations as Eq.(\ref{e30}), with the following matrix of
embedding numbers 
\[
{\tighten
N_{c,g}^{{\rm dim}}=\left(
\begin{array}{ccccccccccccc}
1&0&0&0&0&0&0&0&0&0&0&0&0\\ 
2&1&0&0&0&0&0&0&0&0&0&0&0\\ 
2&0&1&0&0&0&0&0&0&0&0&0&0\\ 
4&4&0&1&0&4&0&0&0&0&0&0&0\\ 
4&0&4&0&1&0&4&0&0&0&0&0&0\\ 
3&2&0&0&0&1&0&0&0&0&0&0&0\\ 
3&0&2&0&0&0&1&0&0&0&0&0&0\\ 
3&1&1&0&0&0&0&1&0&0&0&0&0\\
3&2&1&0&0&0&0&0&1&0&0&0&0\\
5&5&1&1&0&5&0&1&1&1&0&0&0\\
5&2&4&0&1&0&4&2&1&0&1&0&0\\
6&4&2&0&0&3&1&2&0&0&0&1&0\\
7&6&4&1&1&6&4&4&2&2&2&1&1
\end{array}
\right) 
}
\]
The embedding constants for the infinite lattice are
$N_{\infty ,g}^{{\rm dim}}=\frac 14
\left(
4\, 8\, 4\, 1\, 1\, 16\, 4\, 16\, 8\, 8\, 8\, 4\, 4
\right)$,
and the ground state energy per site in the dimer expansion is
\[
E_{{\rm dim}}(J_0,J_1) = J_1\left[ -\frac 38-\frac 3{32}\left(\frac{J_0
}{J_1}\right)^2-\frac 3{128}\left(\frac{J_0}{J_1}\right)^3-0.02295\left(
\frac{J_0}{J_1}\right)^4-
\right.
\]
\begin{equation}
\hspace{30mm}\left.
-0.0213\left(\frac{J_0}{J_1}\right)^5
-0.0240\left(\frac{J_0}{J_1}\right)^6
-0.0205\left(\frac{J_0}{J_1}\right)^7\right]  +...
\label{e60}
\end{equation}

\section{Nonlinear spin waves}

We apply the conventional Holstein-Primakoff formalism, that parametrizes
spins in terms of harmonic oscillators. This approach provides results of a
remarkable precision for the $s=1/2$ Heisenberg antiferromagnet on a square
lattice (see \cite{r050} for review). The CAVO lattice is treated as a
square lattice with nodes containing 4 spins. In the approximation of
noninteracting spin waves, we find 
\begin{equation}
E_{{\rm Neel}}^{{\rm linear}}(J_0,J_1)=
-\left(J_0+\frac 12J_1\right)\left[
s(s+1)-\frac s4\int {\rm Tr}\left\{\sqrt{1-\gamma^2({\bf p},J_0,J_1)}
\right\}\frac{{\rm d}^2p}{\left(2\pi\right)^2}\right]
\label{e70}
\end{equation}
where the integral in ${\bf p}=(p_1,p_2)$ is over the square $2\pi
\times 2\pi $ Brillouin zone. This is analogous to the result for the simple
square lattice except that $\gamma({\bf p},J_0,J_1)$ is now a $4\times 4$
matrix with two exchange parameters 
\[
{\tighten
\gamma({\bf p},J_0,J_1)=\frac 1{2J_0+J_1}\left(
\begin{array}{cccc}
0&J_0&J_1e^{ip_1}&J_0\\ 
J_0&0&J_0&J_1e^{ip_2}\\ 
J_1e^{-ip_1}&J_0&0&J_0\\ 
J_0&J_1e^{-ip_2}&J_0&0
\end{array}
\right) 
}
\]

The next correction of order $O(s^0)$ is the average of quartic terms in the
boson Hamiltonian in the linearly reconstructed ground state. Its direct
evaluation in the Holstein-Primakoff formalism is rather tedious 
(see \cite{r060}), but there exists a way
to simplify the calculation. For bipartite lattices with two equivalent
sub-lattices one can prove 
\[
\left\langle {\bf S}_a{\bf S}_b\right\rangle=\left(s+c_{ab}\right)^2
\]
for any pair of neighbor spins with $c_{ab}$ being a constant of
order $s^0$. In other words, every 
$\left\langle {\bf S}_a{\bf S}_b\right\rangle $ 
is a full square in the nonlinear spin wave approximation.
Therefore, the nonlinear spin wave result may be found by (i) separating the
contributions from different types of nearest neighbors, (ii) completing the
square for each of them and (iii) summing up the results. Thus for the model
(\ref{e10}) we arrive at 
\begin{equation}
E_{{\rm Neel}}(J_0,J_1)=
-\left[ J_0\left(s+\frac 12-C_0\right)^2+\frac 12
J_1\left(s+\frac 12-C_1\right)^2\right]\,,\quad
C_\alpha =
\frac 18\int 
{\rm Tr}\left\{\frac{1-\gamma({\bf p},J_0,J_1)\gamma_\alpha({\bf p})}{
\sqrt{1-\gamma^2({\bf p},J_0,J_1)}}\right\}\frac{{\rm d}^2p}{\left(2\pi
\right)^2}
\label{e80}
\end{equation}
where $\gamma_\alpha({\bf p})$ denotes 
$\gamma_0({\bf p})=\gamma({\bf p},1,0)$ or 
$\gamma_1({\bf p})=\gamma({\bf p},0,1)$. 
For the model under
consideration $s=1/2$ should be substituted.

\section{Discussion}

The results of our estimates for the infinite CAVO lattice are collected
together in Fig.1. In these figures we used the energy of the classical Neel
state $E_{{\rm classical}}(J_0,J_1)=-\left(2J_0+J_1\right) /8$ as a unit
and we plot the rescaled ground state energy $\widetilde{E}=E(J_0,J_1)/E_{{\rm 
classical}}(J_0,J_1)$ as a function of the reduced coupling $\widetilde{J}
=J_1/(J_0+J_1)$ using notations that are similar to those adopted in \cite
{r030}).

For each of the two perturbative expansion we plot the energy in all computed 
orders starting from the 2nd. 
A tentative Pad\'e extrapolation indicates closest
singularities at respectively $J_1/J_0\sim 1.4$ for the plaquette expansion and
$J_1/J_0\sim 1.05$ for the dimer one, but the orders we
reached are not
sufficient to reveal the analytical structure of the expansions. 
However, both expansions appear to converge well in their
dominant region, and we believe that the highest order of perturbation
provides a good estimate for the ground state energy.
Another fact to be noted is that the ground state
energy per site $E_{{\rm rvb}}(J_0,J_0)=-0.5499J_0$ following from
Eq.(\ref{e40}) at the symmetric point $ J_0=J_1$ agrees perfectly with
variatonal Monte Carlo calculations of the singlet state that give
$-0.5510\cdot J_0$ \cite{r070} $-0.5495\cdot J_0$ \cite{r080}.

The precision of the $1/s$ expansion is generally less controlled than that of
ordinary perturbation theories, as the small parameter is not obvious (see
\cite{r050} for discussion).  However, we see that at the symmetric point
($J_1=J_0$) the first correction of order $s$ (linear spin waves) gives 
$\sim45\%$ of the classical Neel energy and the second,  nonlinear correction, 
is $\sim 10\%$ of the first one.  Such a convergence is only a little worse than
in the case of the square lattice where these ratios are $32\%$ and $8\%$,
respectively, and we expect negligible further corrections.

From Fig.1 we see that each of the three states minimizes the energy in some
region of $J_1/J_0$. Namely, the Neel state is stable in the interval
$0.90<J_1/J_0<1.6$, and the plaquette and dimer singlets are stable
correspondingly below and above this region. This is in a perfect quantitative
agreement with the Monte Carlo simulations \cite{r090}, but not with the
approximate treatments of \cite{r100} and \cite{r110}, where the Neel interval
is considerably overestimated. Neither of our three curves for the ground state
energy shows an anomaly in a reasonable vicinity of the intersection point. 
Thus we conjecture the occurrence of discontinuous (first order) transitions as
a function of $J_1/J_0 $, as a result of a direct competition in energy. 
A somewhat similar transition has been reported for a square lattice with
additional frustrating couplings \cite{r115}. In fact, presently
available results obtained by finite temperature simulations or mean-field type
approximations cannot exclude this scenario. It assumes that the gap does not
vanish on the border of the singlet regions, in agreement with recent
perturbative estimates \cite{r120} extrapolated to the point $J_1/J_0=0.9$.  

On the other hand, the fact that we use entirely different approaches for each
region does not allow us to insist on the first order character of the
transitions.  The situation might be sensitive to minor quantitative
changes due to higher order correction, so that the intersecting curves could
turn out to be tangent.

We found it interesting to complete the picture by direct Lanczos type
diagonalization of finite clusters.  The results for four clusters and their
configurations are presented in Fig.2.  Three clusters are cut out from the
infinite lattice with periodic boundary conditions imposed. The cluster of 4
plaquettes is nonfrustrated, whereas those of 5 and 6 plaquettes are both
frustrated.  We have also considered another nonfrustrated cluster of 6
plaquettes designed ``artificially'' as an octahedron with vertices decorated
by plaquettes.  The energies are rather scattered due to small cluster sizes,
but we observe that frustrated clusters tend to follow the singlet perturbation
curves, while non frustrated ones follow the energy of the Neel state.  To
verify our cluster expansion we have computed the perturbation expansion for
the 6 plaquette octahedron (to 5th order in the plaquette phase and to 6th
order in the dimer one). The results shown in Fig.2 suggest that the
perturbation series are very precise in the regions of interest.

The energy spectrum of the frustrated 5 plaquette cluster is rather unusual.
Due to a special symmetry, two singlet states belonging to different
representations intersect near $J_1\approx 1.3\cdot J_0$, while the lowest
triplet state is continuous in $J_1/J_0$. Although this curious example may
be quite special, we consider it as confirming the possibility of first
order transitions on the infinite CAVO lattice.

{\bf Acknowledgment}

We are indebted to Laurent Levy {\em SNCI, Grenoble} for
acquainting us with the $CaV_4O_9$ problem.
S.V.M. thanks {\em Laboratoire de Physique Math\'ematique, Universit\'e 
Montpellier II} for their hospitality and access to computational facilities.

\newpage
\begin{figure}
\epsffile{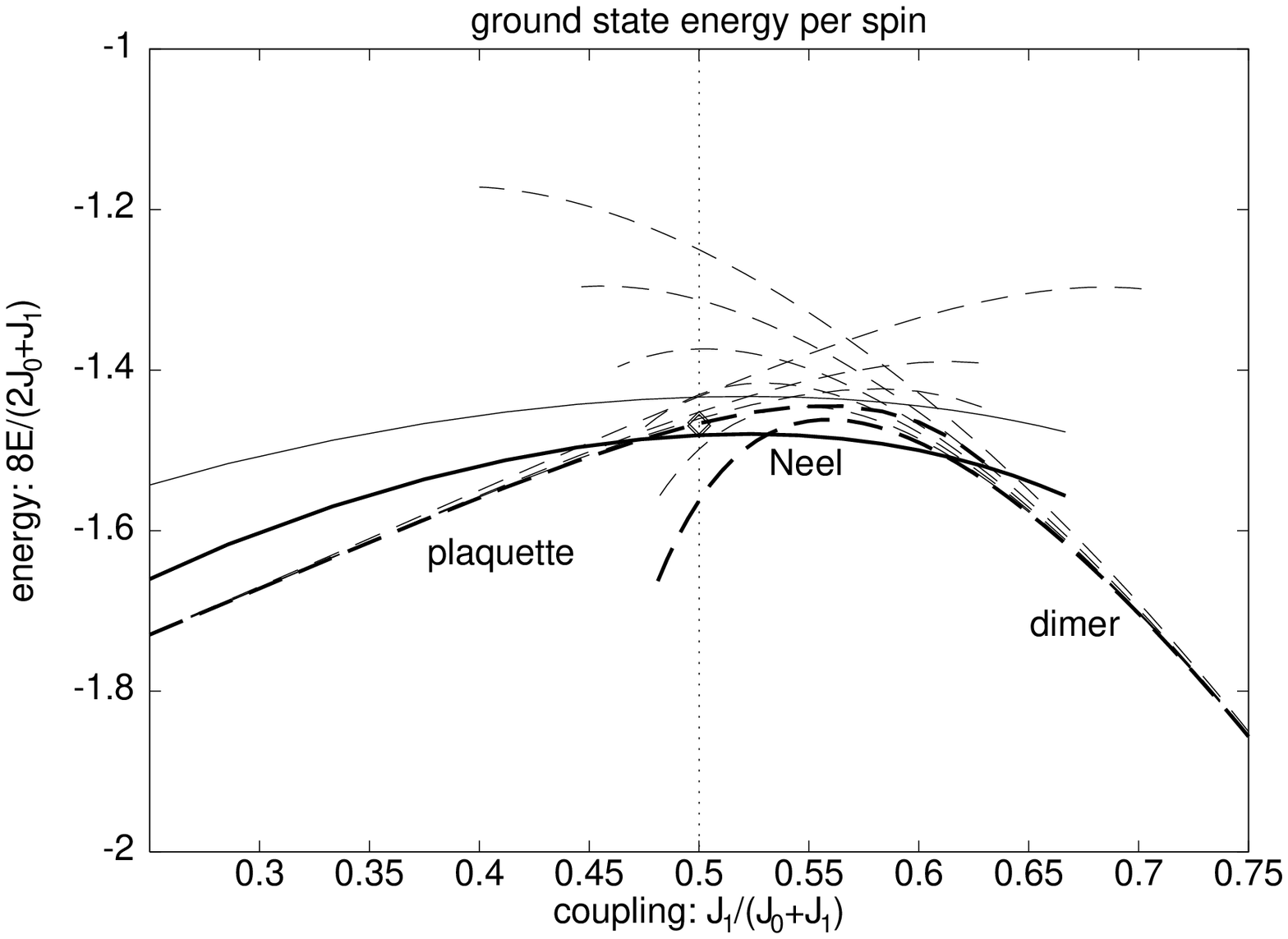}
\vskip 8mm
\caption{
The rescaled ground state energy per spin $\widetilde{E}$ versus the reduced
coupling $\widetilde{J}$. Thin and solid lines represent, respectively, the 
linear and nonlinear spin wave approximations (\protect\ref{e70}) and
(\protect\ref{e80}). Dashed ascending and descending lines represent,
respectively, the perturbation expansions for the plaquette (\protect\ref{e40})
and dimer singlets (\protect\ref{e60}), starting from the 2nd order. The highest
orders (5th for plaquettes and 7th for dimers) are bold-dashed. The two (hardly
distinguishable) diamonds are variational Monte Carlo results 
\protect\cite{r070} and \protect\cite{r080}.
}
\end{figure}
\begin{figure}
\epsffile{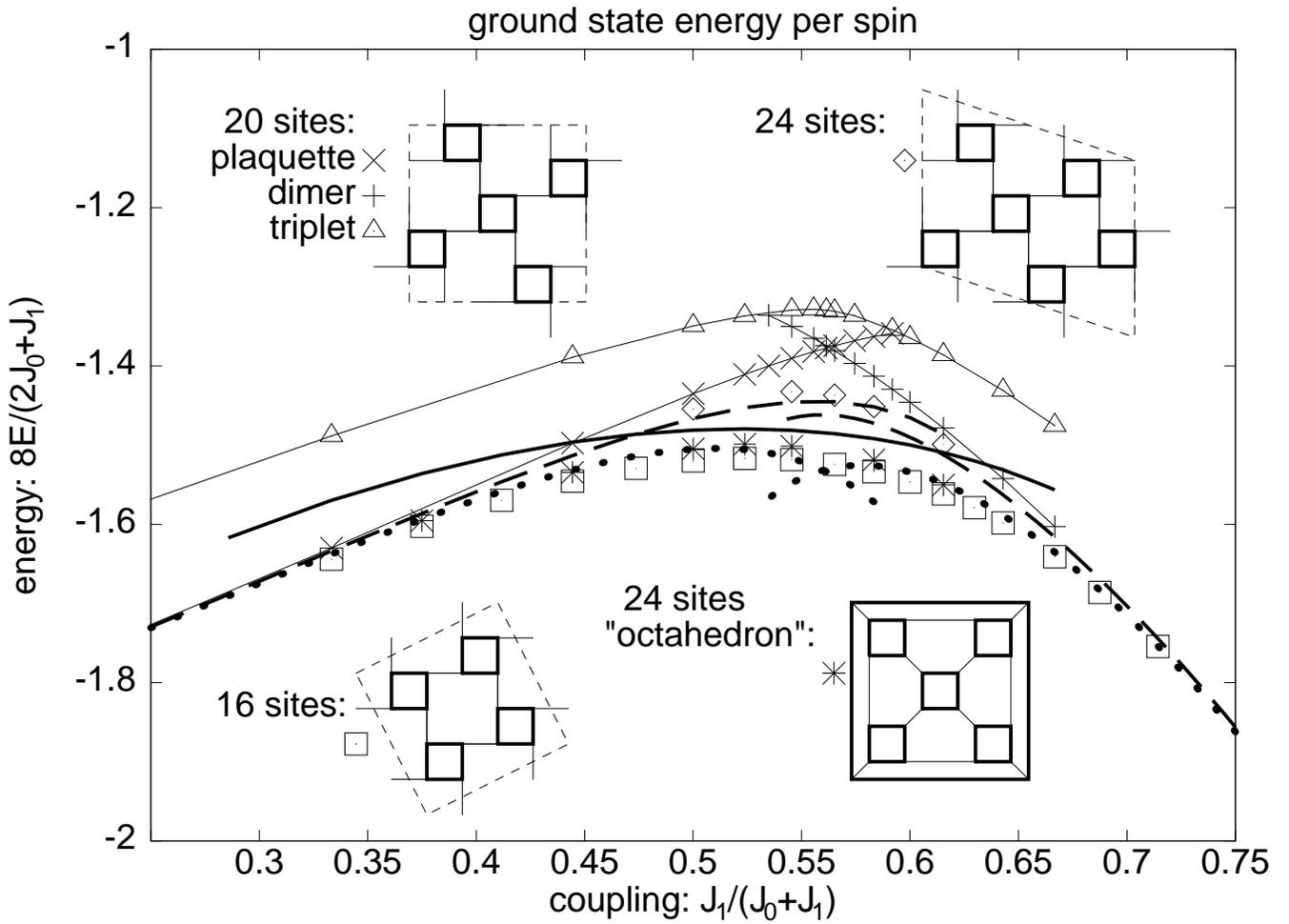}
\vskip 8mm
\caption{
Energies of different finite clusters cut
out from the CAVO lattice and that of an octahedron made up of plaquettes. Bold
solid and dashed lines show the nonlinear spin wave and perturbative results on
the infinite lattice. Dotted lines represent the perturbation expansion for the
24-site octahedron in the plaquette and dimer phase up to, respectively, 5th and
6th order. 
}
\end{figure}

\end{document}